%% file: pt2.tex
\documentclass[11pt]{article}
\usepackage{graphicx}
\usepackage{bm}
\usepackage{amsmath} \usepackage{amsfonts} \usepackage{amssymb} \usepackage{mathtools}
\usepackage{slashed}
\usepackage{mystyle}
\usepackage[sort&compress,numbers]{natbib} 
\newcommand{\mr}[1]{\mathrm{#1}}
\newcommand{\ii}{\mathrm{i}}
\newcommand{\dd}{\mathrm{d}}
\newcommand{\gs}{\mathrm{g}}
\newcommand{\tr}{\mathrm{tr}}
\newcommand{\arctanh}{\mathrm{arctanh}}

\newcommand{\FB}{\mathbf{F}}
\newcommand{\EB}{\mathbf{E}}
\newcommand{\BB}{\mathbf{B}}
\newcommand{\JB}{\mathbf{J}}
\newcommand{\IB}{\mathbf{I}}

\newcommand{\sigmaB}{\bm{\sigma}}
\newcommand{\etaB}{\bm{\eta}}

\newcommand{\kB}{\bm{k}}
\newcommand{\te}{\mathrm{te}}
\newcommand{\tm}{\mathrm{tm}}
\newcommand{\EC}{\mathcal{E}}
\newcommand{\SC}{\mathcal{S}}
\newcommand{\DC}{\mathcal{D}}
\newcommand{\rB}{\bm{r}}
\newcommand{\tB}{\bm{t}}
\newcommand{\Vl}[1]{\stackrel{_\leftarrow}{#1}}
\newcommand{\Vr}[1]{\stackrel{_\rightarrow}{#1}}
\usepackage{xcolor} \definecolor{darkgreen}{rgb}{0,.5,0}
\usepackage[colorlinks,filecolor=blue,citecolor=darkgreen,unicode]{hyperref}

\title{The polarization tensor approach for Casimir effect}

\author{Nail Khusnutdinov$^\dag$ and Natalia Emelianova$\ddag$ \\[3em]
	{\small Centro de Matem\'atica, Computa\c{c}\~{a}o e Cogni\c{c}\~{a}o - Universidade Federal do ABC}\\
	{\small Av. dos Estados, 5001, 09280-560, Santo Andr\'e, SP, Brazil }\\
	{\small $^\dag$nail.khusnutdinov@gmail.com, $^\ddag$natalia.emelianova@ufabc.edu.br}
}

\begin{document}
\maketitle

\begin{abstract}
This paper gives a brief overview of the polarization tensor approach to the Casimir effect. The fundamental principles of this approach are discussed, along with its various applications to both three-dimensional and two-dimensional systems, with a focus on its implications for graphene.
\end{abstract}



\section{Introduction}

Since the famous and concise Casimir paper \cite{Casimir:1948:otabtpcp}, the theory of Casimir effect has been significantly improved and developed. Nowadays, the Casimir effect is a well-established experimental phenomenon (see, for example, book   \cite{Bordag:2009:Ace} and recent review  \cite{Woods:2016:MpCvWi}). Lifshitz \cite{Lifshitz:1956:tmafbs} made the first crucial re-derivation and extension of the theory of the Casimir effect within the framework of Rytov's \cite{Rytov:1987:Psr} theory of electromagnetic fluctuations. The formula derived by Lifshitz depends on the Fresnel reflection coefficients of each boundary. All necessary information regarding the physical properties of solids required for calculating the Casimir free energy and force is encoded in these functions. The primary challenge lies in choosing between models that describe the dielectric properties of materials and the boundary conditions (see Refs.\, \cite{Hartmann:2017:PDMCFPFA,Klimchitskaya:2023:CEIDMTEEWa} for discussions on the Drude and plasma models).

Alternatively, the conductivity can be calculated using Kubo's theory of linear response, a method widely used in the condensed matter community (see, for example, book  \cite{Bruus:2016:Mqtcmpi}). This approach has been applied to materials such as graphene in Ref.\, \cite{Gusynin:2006:TDqgHoc} and to Weyl semimetals in Ref.\, \cite{Rodriguez-Lopez:2014:RCECI}. 

Quantum Field Theory (QFT) offers an alternative method for calculating the dielectric functions of an object and the Casimir effect from fundamental principles. Let us briefly and intuitively discuss the key points of this approach. The Casimir energy can be interpreted as the zero-point energy \cite{Bordag:2009:Ace} which has to be renormalized in a certain way. For example, within the framework of zeta-function regularization, it is represented as the zeta-function of an operator $\DC$ with appropriate  boundary conditions
\begin{equation*}
	\EC(s) = \frac{1}{2}M^{2s} \zeta_{\DC}\left(s - \frac{1}{2}\right),\  \zeta_{\DC}(s) = \sum_\omega \omega^{-2s}.
\end{equation*}
By taking the limit $s\to 0$ with a renormalization procedure, we can determine the Casimir energy \cite{Bordag:2009:Ace}. The parameter  $M$, with dimensions of mass, is arbitrary parameter coming in inevitably. It is used to make correct renormalization procedure.  All information regarding the material and boundary conditions is encoded in the spectrum $\omega$. In most scenarios, this spectrum can be determined  by solving of boundary conditions \cite{Bordag:2009:Ace}. 

The Dyson equation describes an exact Green function in terms of the free Green function and the self-energy $\Sigma$ (see, e.g. book   \cite{Itzykson:1980:Qft}). The poles of the Green function establish an energy spectrum (dispersion relation) and are defined by the self-energy. The self-energy is influenced by all conditions relevant to the problem at hand -- including boundary conditions, material properties, temperature, chemical potential, external fields, etc. The presence of self-energy causes a shift in the spectrum, transforming $\omega \to \omega + \Sigma$ (see, for instance, book  \cite{Bruus:2016:Mqtcmpi}  where the shift due to impurities is described). In the leading order of the fine-structure constant, the self-energy can be expressed in terms of the Polarization Tensor (PT). Consequently, the calculation of the Casimir energy can be framed as follows: the total action is considered as the sum of the electromagnetic action and an effective action incorporating the self-energy. In a (3+1)D scenario with bulk, this total action can be reformulated as the Maxwell action within a dielectric medium, with dielectric properties determined by the self-energy. In the case of infinitely thin surfaces (such as graphene), variations in the total action with respect to the electromagnetic potential yield the Maxwell equations with a current located on the surface.

In our notation, Greek letters represent coordinates in (3+1)D space-time ($\alpha,\beta,\ldots = 0,1,2,3$), while Latin letters from the middle of the alphabet denote coordinates of (2+1)D vectors ($i, j, \ldots = 0, 1, 2$), with spatial components denoted by $a, b, \ldots = 1, 2$.

\section{The polarization tensor}	

The effective action for a single flavor of electron with a one-fermion loop takes the form
\begin{equation}
	S_{\mr{eff}} = \ln \det \slashed{D}_A,
\end{equation}
where the Dirac operator in the presence of an external electromagnetic field is denoted as
\begin{equation*}
	\slashed{D}_A = \ii \left(\slashed{\nabla} + \ii e \slashed{A}\right) - m. 
\end{equation*}

The formal expansion over charge $e$,
\begin{align}
	S_{\mr{eff}} &=  \ln \det \slashed{D}_0 - e\,\tr \left(\slashed{A}\slashed{D}^{-1}_0\right) + \frac{e^2}{2} \tr \left(\slashed{A}\slashed{D}^{-1}_0 \slashed{A} \slashed{D}^{-1}_0 \right) + \ldots\nonumber\\[1ex]
	&= S_{\mr{eff}}^0  + \raisebox{-1em}[0pt][0pt]{\includegraphics[width=4em]{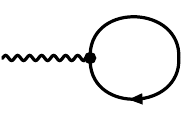}}\ +\  \raisebox{-1em}[0pt][0pt]{\includegraphics[width=5.6em]{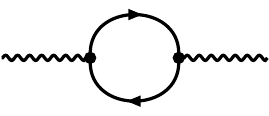}}\ +\ \cdots,
\end{align}
contains an infinite first term which must be removed by a renormalization procedure. The second term corresponds to the tadpole, while the third represents the second-order effective action. 

According to the Furry theorem, \cite{Weinberg:2014:qtfF} the first non-zero term is the second-order term, given by ($D$ is dimension of the space-time) 
\begin{equation}
	S^{(2)}_{\mr{eff}} = \tfrac{1}{2} \iint \dd^D x\ \dd^D y A_\mu(x)\Pi^{\mu\nu}(x;y) A_\nu(y),
\end{equation}
where $\Pi^{\mu\nu}(x;y)$ represents the Polarization Tensor (PT). The Fourier transform of PT reads   
\begin{equation}\label{eq:pmn}
	\Pi^{\mu\nu} (k) = \ii e^2 \int \frac{\dd^D p}{(2\pi)^D} \tr \left(\gamma^\mu  \slashed{D}_0^{-1}(p) \gamma^\nu \slashed{D}_0^{-1}(p-k)\right),
\end{equation}	
where the free Feynman propagator is  
\begin{equation}
	\slashed{D}_0^{-1}(p) = \frac{\slashed{p} +m}{p^2 - m^2+\ii 0}.
\end{equation}

In four-dimensional space-time, the PT \eqref{eq:pmn} is symmetric and gauge invariant under the gauge transformation 
\begin{equation}
	A_\mu(x) \to A_\mu (x) + \nabla_\mu f(x)\ \Leftrightarrow \tilde{A}_\mu(k) \to \tilde{A}_\mu (k) -\ii k_\mu \tilde{f}(k),
\end{equation}
where a tilde indicates a Fourier transform.  This symmetry implies that the PT is a transverse tensor 
\begin{equation}
	\Pi^{\mu\nu} k_\nu = \Pi^{\mu\nu} k_\mu =0.
\end{equation}
The explicit form of the PT in vacuum and its analytical properties can be found in various textbooks, such as Ref.\, \cite{Itzykson:1980:Qft}.

In three dimensions, two distinct representations of gamma matrices exist \cite{Semenoff:1984:CSTA,Hosotani:1993:SbLitgt} which are described by $2\times 2$ matrices Pauli $\sigma_\mu$. The selection of a representation, or specie, is determined \cite{Hosotani:1993:SbLitgt} by the selection number $\eta$ 
\begin{equation}\label{eq:eta}
	\eta = \frac{\ii}{2} \tr\left(\gamma^0\gamma^1\gamma^2\right) = \pm 1.
\end{equation}
These representations take the form\footnote{The Majorana basis see in Ref.\, \cite{Affleck:1982:Issb2d}}
\begin{equation*}
	\gamma^0 =\eta\sigma_3,\ \gamma^1 = \eta \ii\sigma_1,\ 	\gamma^2 = \eta \ii \sigma_2, 
\end{equation*}
obeying the usual anti-commutation relation
\begin{equation*}
	\gamma^\mu \gamma^\nu + \gamma^\nu \gamma^\mu = 2 \gs^{\mu\nu},\ \gs^{\mu\nu} = \mathrm{diag} (1,-1,-1).
\end{equation*}

As a consequence of the non-zero trace of three Dirac matrices, the PT is not symmetric and the action is not  gauge-invariant \cite{Redlich:1984:GNPNTF,Redlich:1984:Pvgnegfatd}. The Pauli-Villars renormalization procedure restores gauge invariance but introduces a parity anomaly in the form of a topological term in the effective action. Upon taking the limit of infinite Pauli-Villars masses $M_{\mr{PV}} \to \infty$ the Chern--Simons contribution  survives in the action
\begin{equation}
	S_{\mr{top}} = -\frac{\ii\eta}{8\pi} \mathrm{sign}\, (m) \int \dd^3 x \varepsilon^{ijl} A_i \partial_j A_l.
\end{equation}
The PT can be decomposed into symmetric and anti-symmetric components, $\Pi^{\mu\nu} = \Pi^{\mu\nu}_{\mr{s}} + \Pi^{\mu\nu}_{\mr{a}}$, where 
\begin{equation*}
	\Pi^{\mu\nu}_{\mr{s}} = \left(\gs^{\mu\nu} - \frac{k^\mu k^\nu}{k^2}\right) \Pi_{\mr{s}},\ \Pi^{\mu\nu}_{\mr{a}} = \eta \epsilon^{\mu\nu\alpha} k_\alpha \Pi_{\mr{a}},
\end{equation*}
ensuring the overall gauge invariance of the PT, $\Pi^{\mu\nu} k_\nu = 0$.  

Studies such as Refs.\, \cite{Semenoff:1984:CSTA,Redlich:1984:Pvgnegfatd} demonstrate that different species yield the same Chern-Simons term with opposite signs ($S_{\mr{top}}  \sim \eta$). Consequently, the combined contribution of both species to the polarization tensor results in a symmetric and gauge-invariant tensor. By representing both species in a  $4\times 4$ reducible representation of gamma matrices, expressed as
\begin{equation*}
	\gamma^0 = 
	\begin{pmatrix}
		\sigma_3 & 0\\
		0 & -\sigma_3
	\end{pmatrix},\ 
	\gamma^1 = 
	\begin{pmatrix}
		\ii \sigma_1 & 0\\
		0 & -\ii \sigma_1
	\end{pmatrix},\ 
	\gamma^2 = 
	\begin{pmatrix}
		\ii \sigma_2 & 0\\
		0 & -\ii \sigma_2
	\end{pmatrix},
\end{equation*}
which is a direct sum of two irreducible representations, the convergence and gauge invariance of the PT in the context of graphene were recently verified in Ref.\, \cite{Bordag:2024:Cptstd}. 

\section{Condensed matter applications}

In the (3+1)D case, we consider the total action \cite{Bordag:2021:BcCiDm} 
\begin{equation}
	S_{\mr{T}} = S_{\mr{M}} + 	S^{(2)}_{\mr{eff}},
\end{equation} 
where 
\begin{equation}
	S_{\mr{M}} = \tfrac{1}{4} \int \frac{\dd^4 k}{(2\pi)^4} \left[\varepsilon_0 \EB (-k)\cdot \EB (k) - \mu^{-1}_0 \BB (-k)\cdot \BB (k)\right],
\end{equation}
represents the action of the electromagnetic field in a medium with bare dielectric permittivity $\varepsilon_0$ and magnetic permeability $\mu_0$. This total action can be represented as the action of the electromagnetic field in the medium with the dielectric function \cite{Lindhard:1954:pgcp} 
\begin{equation*}
	\varepsilon_{ab} = \left(\delta_{ab} - \frac{k_a k_b}{\kB^2}\right) \varepsilon_t + \frac{k_a k_b}{\kB^2} \varepsilon_l,
\end{equation*} 
where the scalars $\varepsilon_t,\varepsilon_l$ are expressed \cite{Bordag:2021:BcCiDm} in terms of the PT 
\begin{equation}
	\varepsilon_l = \varepsilon_0 + \frac{\Pi^{00}}{\kB^2},\ \varepsilon_t = \varepsilon_0 - \frac{\kB^2}{k_0^2} \left(\frac{1}{\mu_0} -1\right) - \frac{1}{2 k_0^2} \left(\frac{k_0^2 - \kB^2}{\kB^2} \Pi^{00} + \Pi^\mu_\mu\right).
\end{equation}

Consider a 2-dimensional conductive plane in vacuum perpendicular to the $x^3$ axis at the point $x^3 =d$ in 3-dimensional space. The total action is a sum $S_{\mr{T}} = S_{\mr{M}} + 	S^{(2)}_{\mr{eff}}$ of the (3+1)D action of the Maxwell vacuum field 
\begin{equation*}
	S_{\mr{M}} = -\tfrac{1}{4} \int \dd^4 x F_{\mu\nu} F^{\mu\nu},
\end{equation*}
and (2+1)D effective action 
\begin{equation}
	S^{(2)}_{\mr{eff}} = \tfrac{1}{2} \iint \dd^3 x\ \dd^3 y A_i(x)\Pi^{ij}(x;y) A_j(y).
\end{equation}

The Maxwell equations are obtained by varying the total action with respect to the electromagnetic potential (setting $\Pi^{3\mu} = \Pi^{\mu3} = 0$)
\begin{equation}
	F^{\mu\nu}_{\ ,\nu} = - \delta(x^3 - d) \int \dd^3 y \Pi^{\mu\nu}(x-y) A_\nu(y) = - 4\pi J^\mu,
\end{equation}
with a current $J^\mu$ on the plane. Integration near the plane yields the boundary conditions
\begin{equation}\label{eq:Jgen}
	[\BB] \times \bm{n} = 4\pi \JB,\ [\EB]\cdot \bm{n} = - 4\pi \rho,\ \JB = \sigmaB \EB,
\end{equation}
where all relations are considered on the plane $x^3 =d$ and $[f(x^3)] = f(x^3 - 0) - f(x^3+0)$ denotes the  jump of function on the plane. Here $\bm{n} = (0,0,1)$ is a vector perpendicular to plane, and 
\begin{equation}
 \sigma^{ab} = \frac{\Pi^{ab}}{\ii \omega}, \ \rho = \frac{\Pi^{0a} E_a}{\ii \omega}.
\end{equation}
In the Weyl gauge, $A_0=0$,  the Ohm's law \eqref{eq:Jgen} can be rewritten as $J^a = \Pi^{ab} A_b$ where the tensor $\Pi^{ab}$ is known in microscopic condensed matter physics and plasma physics as the response tensor  \cite{Melrose:2008:Qpup}. The conductivity of the plane can be determined by calculating the corresponding PT.

This approach finds a natural application in 2D materials, particularly in graphene. \cite{Novoselov:2004:Efeatcf} For energies below $3$ eV, the behavior of electrons in graphene can be described by the Dirac equation with Fermi velocity $v_F \sim 1/300$ and a mass gap $m < 1$ eV.  Formally, in terms of the new matrices $\tilde{\gamma}^0 = \gamma^0, \tilde{\gamma}^1 = v_F\gamma^1, \tilde{\gamma}^2 = v_F \gamma^2$ we have the same Dirac equation but in the (2+1)D space-time with the  metric $\left[\tilde{\gs}^{il}\right] = \mathrm{diag}(1,-v_F^2,-v_F^2)$.  One has two species of fermions and two sub-lattices A and B.

The total action which describes graphene is not Lorentz invariant. Making the Lorentz transform of action  \cite{Antezza:2024:nCLflmg}  with 3-velocity $[u^l] = (u^0,u^1,u^2)$
\begin{equation*}
	S = - \tfrac{1}{4} \int  \dd^4x \FB^2 + \int \dd^3 x \bar{\psi} \left[\tilde{\gamma}^l (\ii \partial_l - e A_l) -m \right]\psi |_{z=a},
\end{equation*}
along the graphene sheet one obtains the action 
	\begin{equation*}
		S' = - \tfrac{1}{4} \int  \dd^4x' \FB'^2 + \int \dd^3 x' \bar{\psi}' \left[\tilde{\Gamma}^l (\ii \partial_{l'} - e A'_l) -m \right]\psi'|_{z=a},
	\end{equation*}
and $S' \not = S$. Here 
\begin{equation*}
		\tilde{\Gamma}^{l}  = v_F \gamma^l + u^l (1-v_F) \slashed{u} \not = \tilde{\gamma}^l,\ [u^n] = (u^0,u^1,u^2).
\end{equation*}

The PT generated by a single planar quasi-relativistic fermions at zero temperature and chemical potential was found in Ref.\, \cite{Bordag:2009:CipcgdDm}. It has the following form
\begin{equation}
	\Pi^{il} = \frac{e^2 }{v_F^2} \left[m\psi\left( \tilde{\gs}^{il} - \frac{k^i k^l}{k^2}\right) + \ii v_F^2 \phi \varepsilon^{ilj} k_j\right],
\end{equation}
where
\begin{align*}
	\psi  &=  (n_+ + n_-)  \left[1 - \left(\frac{k}{2m} + \frac{2m}{k}\right) \arctanh \left( \frac{k}{2m}\right)\right],\\
	\phi  &= (n_+ - n_-) \left[\frac{2m}{k}\arctanh \left( \frac{k}{2m}\right) - 1 \right],
\end{align*}
and $n_\pm$ are numbers of fermion's species in each sublattice. Here we are working in (2+1)D space-time with metric $\left[\tilde{\gs}^{il}\right] = \mathrm{diag}(1,-v_F^2,-v_F^2)$ and therefore $k^i = \tilde{\gs}^{il}k_l$ and $k^2 = k^i k_i = k_0^2 - v_F^2 \kB^2$.  For graphene $n_\pm=2$ and the parity anomaly has no contribution to the PT. 

The conductivity tensor can be diagonalized to obtain eigenvalues
\begin{equation*}
	\sigma_\tm = \frac{\ii \omega}{\kB^2} \Pi_{00},\  \sigma_\te = \frac{1}{\ii \omega} \left(\Pi^n_n + \frac{\omega^2 - \kB^2}{\kB^2} \Pi_{00}\right),
\end{equation*}
corresponding to the conductivity of the transverse magnetic and transverse electric polarization.

When considering an infinitely thin plane such as graphene approaching the zero thickness film limit, $L\to 0$,  a problem arises \cite{Klimchitskaya:2014:ClCitfag}  in calculating the Casimir energy and force using the well-known Lifshitz formula. The issue lies in the fact that the reflection coefficients tend towards zero in this limit, causing the Casimir free energy and pressure to also approach zero.  Utilizing a tensorial form of conductivity requires the consideration of reflection Fresnel matrices rather than scalars. One method to determine the reflection coefficients is to use the PT approach described here and the scattering matrix approach \cite{Jaekel:1991:Cfptm} to calculate the Casimir energy. In a previous study \cite{Klimchitskaya:2013:vWCitgs},  the Casimir energy between two graphene sheets was computed using the PT approach. 

\begin{figure}[t]
	\centerline{\includegraphics[width=.5\linewidth]{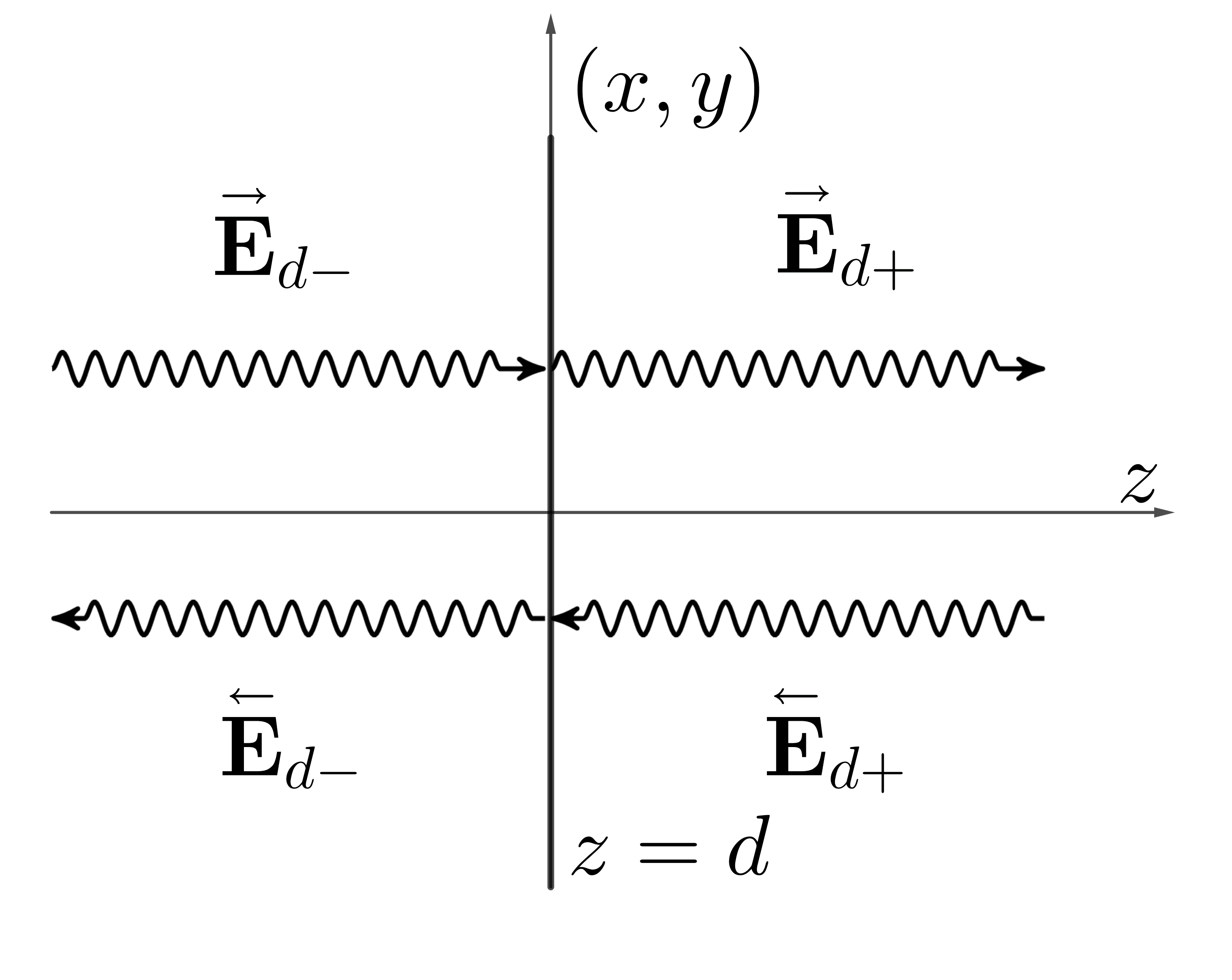}}
		\caption{The plane is perpendicular to axis $z$ at point $d$. Here $\Vr{\EB} \sim e^{+\ii k_3 z}$ and $\Vl{\EB} \sim e^{-\ii k_3 z}$.}\label{fig:S}
\end{figure}

In the framework of the scattering matrix approach, the Casimir energy between two planes separated by a distance $a$ is given by the following expression (see, for example, Ref.\, \cite{Fialkovsky:2018:Qcrcs}): 
\begin{equation}
	\EC=\frac 1{4\pi} \int \frac{\dd^2 k}{(2\pi)^2}\int_{-\infty}^{\infty}\dd\xi\, \ln \det \left(1 - e^{-2a\sqrt{\xi^2 + \kB^2}}\rB'_{\mr{I}} \rB_{\mr{II}}\right), 
\end{equation}
where the reflection matrices are evaluated at imaginary frequency $\omega = \ii \xi$. The scattering matrix for a plane positioned perpendicular to $z=d$ is defined by a specific relation
\begin{equation*}
	\begin{pmatrix}
		\Vl{\EB}_{d-}\\
		\Vr{\EB}_{d+}
	\end{pmatrix} =
	\SC_d \begin{pmatrix}
		\Vr{\EB}_{d-}\\
		\Vl{\EB}_{d+}
	\end{pmatrix},\ \SC_d 
	= \begin{pmatrix}
		\rB_d & \tB'_d\\
		\tB_d & \rB'_d
	\end{pmatrix},
\end{equation*}
with notation for the electromagnetic fields illustrated in Fig.\,\ref{fig:S}.
The explicit form of the reflection matrices is determined from boundary conditions on the plane \eqref{eq:Jgen}. The general form of the reflection coefficients was derived in Ref.\, \cite{Fialkovsky:2018:Qcrcs},
\begin{equation*}
	\rB = \rB' = -\frac{\omega^2 \etaB - \kB \otimes \kB\etaB + \IB \omega k_3 \det\etaB}{\omega^2 \tr \etaB - \kB\kB\etaB + \omega k_3 (1 + \det\etaB) },\ \tB = \IB + \rB. 
\end{equation*}
where\footnote{In the SI units one has $\etaB = \sigmaB/2$} $\etaB = 2\pi \sigmaB, k_3 = \sqrt{\omega^2 - \kB^2}$ and $ \kB\kB\etaB = k_a k_b \eta^{ab}$. 

This approach can be extended to several graphene-specific scenarios. These include
\begin{itemize}
	\item Non-zero temperature \cite{Fialkovsky:2011:FCeg,Bordag:2015:Qftdrg} (Matsubara frequencies): \vspace*{-1ex}
	\begin{equation*}
		\int \dd k_0 f(k_0) \Rightarrow 2\pi\ii T \sum_{k=-\infty}^\infty f(2\pi\ii T (k+1/2)),\ p_0 \to 2\pi \ii n T 
	\end{equation*}
	\item Chemical potential (doped graphene) \cite{Fialkovsky:2011:FCeg,Bordag:2016:ECedg}: $\partial_0 \to \partial_0 - \ii \mu$, $k_0 \to k_0 +\mu$
	\item Impurities  \cite{Fialkovsky:2012:Frg,Khusnutdinov:2024:IgtiCi}: $k_0 \to k_0 + \ii \Gamma \mathrm{sign}\, k_0$
	\item External magnetic field $\BB$: $\partial_k \to \partial_k  + \ii e A_k$ (Faraday rotation in graphene) \cite{Fialkovsky:2012:Frg}
	\item Strained graphene  \cite{Bordag:2017:Cisg}: $v_F \to v_F^{ab} = v_F \left[\delta^{ab} - \frac{\beta}{4} \left(2 u^{ab} + \delta^{ab} u^c_c\right)\right]$
	\item Lateral motion of graphene sheets  \cite{Antezza:2024:nCLflmg,Khusnutdinov:2024:NCFLMPIC}   -- normal Casimir force and friction 
	\item Stack of graphene sheets \cite{Emelianova:2023:Cesgs}
	\item The Casimir torque of an anisotropic molecule in proximity to graphene sheets \cite{Antezza:2020:CftamccpteC}
\end{itemize}

\section{Conclusion} 

This paper provides a brief overview of the application of the polarization tensor (PT) to the Casimir effect. The PT represents the leading non-zero term in the self-energy expansion over the fine-structure constant, effectively describing the frequency spectrum incorporating the physical conditions of the system.  The approach is based on the consideration of the total action of the system which consists of the Maxwell electromagnetic part and the effective action, which takes into account boundary conditions, shape, temperature, etc. Several applications of the PT to graphene are explored. 

\section*{Acknowledgments}

NK was supported in part by the grants 2022/08771-5, 2021/10128-0 of S\~ao Paulo Research Foundation (FAPESP). One of us (NK) is grateful to M. Bordag, C. Henkel, V. Mostepanenko  and D. Vassilevich for fruitful discussions.  

\bibliographystyle{apsrev4-1}
\input{pt2bbl}

\end{document}

%% file: pt2bbl.tex
%